\documentstyle[twocolumn,prl,epsfig,aps]{revtex}
\begin{document}
%
\twocolumn[
\hsize\textwidth\columnwidth\hsize\csname@twocolumnfalse\endcsname
%
\title{On the athermal character of structural phase transitions}
\author{ Francisco  J.  P\'erez-Reche, Eduard Vives, 
Llu\'{\i}s Ma\~nosa, and 
Antoni Planes}
\address{Departament d'Estructura i Constituents de la Mat\`eria, \\
Facultat de F\'\i sica, \\
Universitat de Barcelona. \\
Diagonal, 647, E-08028 Barcelona, Catalonia}
\maketitle
\begin{abstract}
The significance  of thermal fluctuations on  nucleation in structural
first-order phase  transitions has been examined.   The prototype case
of  martensitic transitions  has been  experimentally  investigated by
means of  acoustic emission techniques.   We propose a model  based on
the  mean   first-passage  time   to  account  for   the  experimental
observations.  Our study provides a unified framework to establish the
conditions for isothermal and athermal transitions to be observed.
\end{abstract}
]
%
Fluctuations  are considered as  essential for  a phase  transition to
take  place.  However,  at moderate  temperatures many  solids undergo
first-order phase  transitions which  are regarded as  being athermal.
The  kinetics  of  these  transitions  is  not  dominated  by  thermal
fluctuations  and, consequently, they  can only  take place  under the
change  of an external  parameter (stress,  magnetic field,temperature
...)  which modifies  the free energy difference between  high and low
symmetry phases.  This behaviour is in contrast to thermally activated
transitions for which the relaxation from a metastable state can occur
at constant external conditions  due to thermal fluctuations.  Typical
examples  of athermal  transitions are  found within  the  families of
magnetic  and structural  phase transitions  which  involve long-range
interactions  (dipolar, elastic,  \dots).  The  Martensitic Transition
(MT) undergone by many solids is  an interesting case that will be the
prototype  system experimentally investigated  in the  present letter.
It  is a diffusionless  first-order transition  \cite{Otsuka98} mainly
caused by a shear  mechanism \cite{Nishiyama78}.  Often MT is preceded
by interesting premonitory effects such as softening of long and short
wavelength  acoustic vibrational  modes.  The  anharmonic  coupling of
these  modes  is  at  the  origin  of  the  instability  of  the  high
temperature phase \cite{Planes01}.
 
In general, when  a system is externally driven  through a first-order
phase transition,  the response to  the external control  parameter is
determined by  the characteristics  of the energy  barriers separating
the two  phases. When thermal fluctuations are  not relevant (athermal
transition), the  system remains in  a given configuration as  long as
the state corresponds to a free energy local minimum. When driven, the
system  jumps  towards  a  different  configuration,  once  the  local
stability  limit is  reached.   The  path followed  by  the system  is
strongly influenced by the  existence of disorder (dislocations, grain
boundaries,   vacancies,  local  composition,   atomic  configuration,
etc...), which  controls the  actual distribution of  energy barriers.
As  the system  evolves, it  passes through  a sequence  of metastable
states.  The kinetics is  characterised by jumps (avalanches) from one
metastable state to another which occur with a certain relaxation time
$\tau_{ava}$.  In  a number of  cases this time  has been found  to be
power law  distributed \cite{Vives94}.  The  jerky response for  MT is
related to sudden  changes in the local strain field  which are at the
origin of  the emission of acoustic  waves in the range  from $kHz$ to
$MHz$  \cite{Yu87}.  This  effect is  the so-called  acoustic emission
(A.E.).   Similar  crackling behaviour  is  observed in  ferromagnetic
systems  (Barkhausen  noise)  \cite{Vergne81}, capillary  condensation
systems \cite{Lilly96} and others \cite{nature}.  In the athermal case
the path followed by the system  can be reproduced from cycle to cycle
provided that disorder does not evolve \cite{Sethna93}.

Traditionally MT have been  regarded as athermal \cite{Cao90}, however
very  recent  experimental  observations  of  the  occurrence  of  the
transition at constant temperature question this athermal character in
shape-memory alloys \cite{Aspelmeyer99}.   The present letter is aimed
at  providing a  simple explanation  for the  observation  of apparent
athermal  behaviour and  describing  the possibility  of a  cross-over
between  such  a  behaviour  and  cases where  thermal  activation  is
evident.   The  problem  is   intimately  related  to  the  nucleation
mechanism. Classical nucleation theories  have proven to be inadequate
for MT and heterogeneous nucleation models have been proposed, most of
them neglecting the effect  of themal fluctuations \cite{Vul93}.  This
problem is central for the understanding of MT and remains an issue of
active debate \cite{Vul93,Cao90}.   We have investigated two prototype
shape-memory  materials   undergoing  a  MT  by   detecting  the  A.E.
generated during the transition.   The interest in the A.E.  technique
lies  in  the  fact  that   it  is  highly  sensitive  to  very  small
microstructural changes taking place in the studied sample.  It is the
best suited technique to detect the nucleation and growth of any small
domain of the low temperature phase.
   
We   have    selected   two   single    crystals   with   compositions
Cu$_{68.4}$Al$_{27.8}$Ni$_{3.8}$                                    and
Cu$_{68.0}$Zn$_{16.0}$Al$_{16.0}$.  They underwent, respectively, a MT
from the $\beta$ (ordered-$bcc$)  towards a $\gamma '$ (hexagonal) and
a $\beta'$ (orthorhombic) martensites.   The samples were annealed for
30 min at 1173 K and  then rapidly cooled down to room temperature and
further annealed  for several  days.  A large  number of  cycles (more
than 50)  were done  in order to  have a  reproducible transformation.
A.E  signals  were detected  by  a  resonant piezoelectric  transducer
acoustically  coupled to  the top  flat  surface of  the sample.   The
bottom   flat  surface  was   mounted  onto   a  Cu-block   which  was
cooled/heated  by a  Peltier  element.  In  order  to detect  possible
isothermal  effects it is  of major  importance to  avoid uncontrolled
temperature   oscillations  and/or   drifts.    A  computer-controlled
feed-back  procedure   provided  temperature  control   with  relative
oscillations smaller than  0.01 \%.  The absolute value  of the sample
temperature  \cite{Calvet63}  is  known  within  $\pm  0.5$  K.   A.E.
signals  were  counted using  a  frequency counter  with  a  1 s  gate
\cite{porta}.  Background  noise is inherent to the  detection of A.E;
this noise  was quantified by performing experiments  at a temperature
of more than  100 K above $M_s$.  The background count  rate is in the
range  of 10  Hz.  This  renders a  signal-to-noise ratio  better than
10$^3$ during the M.T.

\begin{figure}[th]
\begin{center}
\epsfig{file =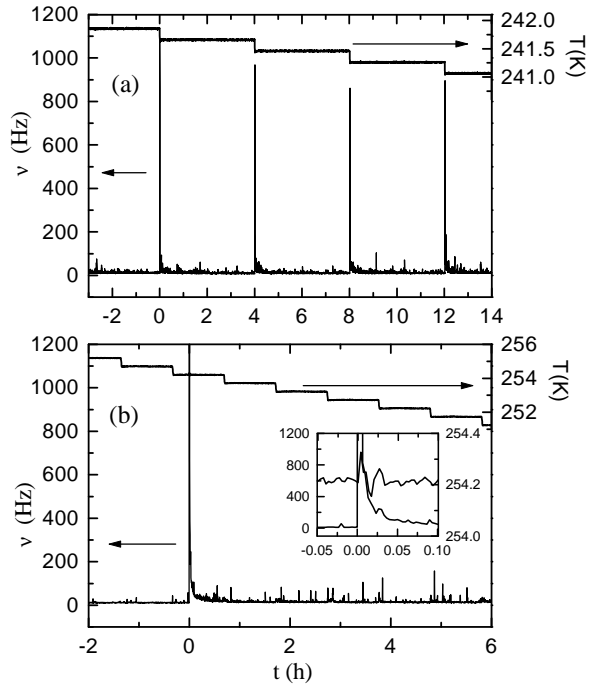, width=8cm}
\end{center}
\caption{Typical  examples  of  the  A.E.   recorded  during  stepwise
isothermal  runs  for  (a)  Cu-Zn-Al  and (b)  Cu-Al-Ni.   Notice  the
difference  in the  temperature  scale (right  axis)  between the  two
plots.  The  inset in  figure (b)  shows a detailed  view of  the A.E.
recorded isothermically.  The large  fluctuation in the temperature is
caused   by   the  latent   heat   released   during  the   isothermal
transformation of a martensitic domain.}
\label{FIG1}
\end{figure}

Two  complementary  experimental  procedures  were followed  for  each
studied sample.   In the first case,  the sample was  cooled down from
room  temperature in  a stepwise  manner.  Each  step consisted  of an
isothermal  plateau lasting  a  time  $t$, followed  by  a cooling  at
$\dot{T}=  1$  K/min  down  to  the  next  plateau.   The  temperature
difference  between consecutive  plateaux was  $\Delta T$  and several
values for  $t$ and $\Delta  T$ were tested.  This  procedure provided
very  fine  tuning  for  the  detection  of  any  possible  isothermal
nucleation and growth of martensite.  A typical example of the results
obtained   is  shown   in  Fig.    \ref{FIG1}.   For   Cu-Zn-Al  (Fig.
\ref{FIG1}a), no trace of  A.E.  was detected under isothermal holding
conditions,  for any value  of $\Delta  T$ (values  as low  as $\Delta
T=0.2$ K  were investigated).  Moreover,  below a  certain temperature,
A.E.  resumes  each time the temperature is  decreased.  The behaviour
found for Cu-Al-Ni (Fig. \ref{FIG1}b)  turns out to be different.  For
isothermal plateaux  separated by  more than 0.8  K, the  behaviour is
similar to  that reported for Cu-Zn-Al.  However,  for smaller $\Delta
T$, A.E.  under isothermal conditions is detected.  A detailed view is
presented in  the inset.  The  overshoot in the temperature  caused by
heat release is clear evidence  that the detected acoustic signals are
indeed  generated  by  the  transformation of  a  martensitic  domain.
Furthermore, it is worth noting  that in this case A.E.  continues for
many  minutes   during  the  isothermal  plateau   above  the  average
background noise.  From these stepwise experiments, it can be inferred
that thermal fluctuations  do trigger the M.T. for  Cu-Al-Ni, when the
system is close enough to  the stability limit.  In contrast, it seems
that thermal  fluctuations are not  able to trigger the  transition in
Cu-Zn-Al  even  in  the case  when  the  system  is in  the  two-phase
co-existence region.

\begin{figure}[th]
\begin{center}
\epsfig{file =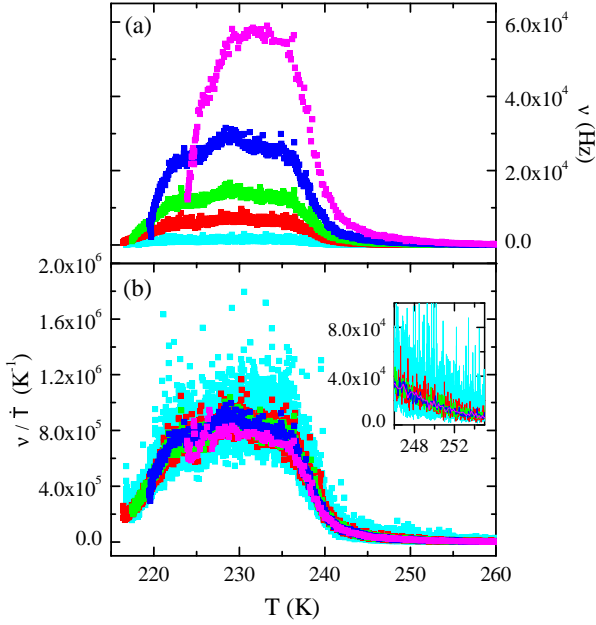, width=8cm}
\end{center}
\caption{(a) A.E. recorded during continuous cooling runs for Cu-Zn-Al
at cooling rates of $\dot T$=  5 K/min (pink), $\dot T$= 2 K/min (dark
blue), $\dot T$= 1 K/min (green), $\dot T$= 0.5 K/min (red), $\dot T$=
0.1  K/min (blue).   (b) Scaled  acoustic  activity as  a function  of
temperature, showing  the scaling of the different  curves.  The inset
shows an enlarged view of the tail of the curves.}
\label{FIG2}
\end{figure}

\begin{figure}[th]
\begin{center}
\epsfig{file =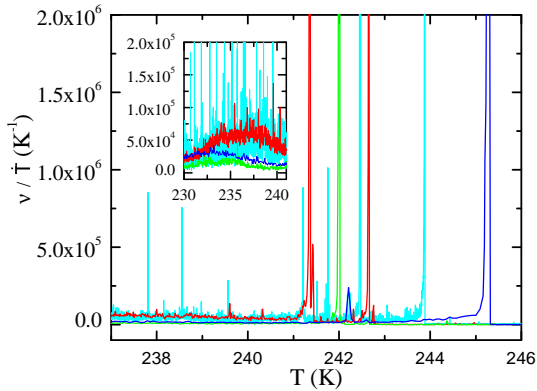, width=8cm}
\end{center}
\caption{Scaled  A.E.   recorded during  continuous  cooling runs  for
Cu-Al-Ni at cooling rates of $\dot  T$= 2 K/min (dark blue), $\dot T$=
1 K/min  (green), $\dot T$=  0.5 K/min (red)  and $\dot T$=  0.1 K/min
(blue). The inset shows a detail of the low temperature region.}
\label{FIG3}
\end{figure}

\begin{figure}[th]
\begin{center}
\epsfig{file =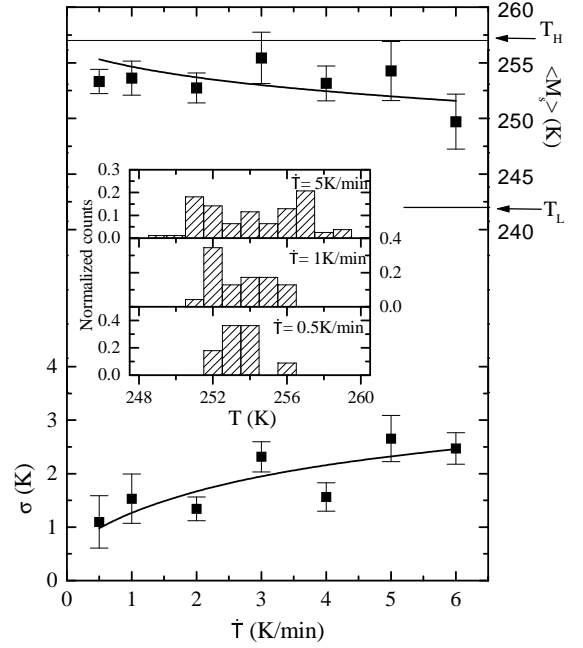, width=8cm}
\end{center}
\caption{Average  value of  the transition  temperatures  $\langle M_s
\rangle$ and its standard deviations $\sigma$ (symbols), as a function
of the cooling rate $\dot T$. Thick continuous lines correspond to the
fits  of the three-parameter model  proposed in  the text.   The thin
lines indicate the fitted values  of $T_H$ and $T_L$.  The inset shows
examples of histograms of $M_s$ distributions.}
\label{FIG4}
\end{figure}

The  second  procedure consists  of  measuring  A.E. under  continuous
cooling at different rates $\dot T$. Results for Cu-Zn-Al are shown in
Figure \ref{FIG2}a.   We propose  that, for athermal  transitions, the
absence  of kinetic  effects will  result in  a scaling  of  the ratio
between the acoustic activity  and the cooling rate ($\nu$/$\dot{T}$).
We have evaluated the actual  cooling rate at each temperature and the
results for  $\nu$/$\dot{T}$ are plotted as a  function of temperature
in  Fig. \ref{FIG2}b.   Excellent  scaling is  observed  over all  the
temperature range and a magnified  view of the high temperature region
is shown in the inset.  The larger scatter for smaller $\dot T$ is due
to the  propagation of the  error in the temperature  measurement (the
relative error in $\nu/{\dot T}$ is proportional to $1/{\dot T}$). The
results obtained  for Cu-Al-Ni are presented in  Fig.  \ref{FIG3}.  No
scaling is  observed over all  the temperature domain  thus indicating
that the transition does not have an athermal character for this alloy
system.  Instead,  the transition is  not reproducible and  the values
for   $M_s$   exhibit  a   stochastic   character.   The   probability
distribution  of  $M_s$ can  be  characterized  by  its average  value
$\langle M_s  \rangle$ and  its standard deviation  $\sigma =  \left (
\langle  M_s^2 \rangle  - \langle  M_s \rangle^2  \right)^{1/2}$.  The
behaviour of  these two quantities as  a function of  $\dot{T}$ can be
estimated by  performing a  statistical analysis of  $M_s$ data  for a
large  number  of  temperature  loops.   Results  are  shown  in  Fig.
\ref{FIG4} (solid  symbols).  The total number of  cycles analysed for
each value  of $\dot T$ ranges  from 15 for  $\dot{T}=0.5$K/min to 150
for $\dot{T}=6$K/min.


A proper  statistical analysis requires the  detection and suppression
of  any possible  systematic  deviation of  the  data or  even a  fake
dependence on  $\dot T$ which  may arise from ageing  phenomena during
the experiment.   To prevent  these undesired effects,  the experiment
was performed  as a series  of temperature loops  ($220$K $\rightarrow
340$K$\rightarrow 220$K) and each series corresponded to a given value
of  $\dot T$.   The series  are separated  by low  temperature waiting
periods randomly selected within the  range from 1 to 1000 hours.  The
studied  values for  $\dot T$  within the  sequence are  also randomly
chosen, and each value appears in the sequence several times.  A small
drift of $\langle M_s \rangle$  ($< 1.4 $K/month) with ageing time was
detected and  corrected from the  data displayed in  Fig.  \ref{FIG4}.
The standard deviation $\sigma$ was found to be insensitive to ageing.
The  inset  illustrates  the  actual  histograms of  $M_s$  for  three
different values of $\dot T$.

The behaviour of  $\langle M_s \rangle$ and $\sigma$  can be explained
by an adequate theory based  on the analysis of the mean first-passage
time  \cite{Perez00} between  the $\beta$  and the  martensitic phase.
Here we briefly summarise  this theoretical approach.  The probability
$P(t)$ for the system to remain in the metastable phase after time $t$
is given by: $dP(t)/dt = - \lambda(t) P(t)$, where $\lambda(t)$ is the
transition  probability per  unit time.   As for  certain  systems the
transition seems to  be athermal, while it is not  in others, our main
assumption  in  reproducing  the  experimental  observations  is  that
thermal  fluctuations  are   only  active  within  two  characteristic
limiting  temperatures: when  $T>T_H$, $\lambda=0$  (no  transition is
possible), and  when $T<T_L$, $\lambda=\infty$  (the transition occurs
with  absolute certainty).   This  assumption leads  to the  following
expressions  for  the  average  temperature  transition  $\langle  M_s
\rangle$ and its fluctuations $\sigma^2$:
\begin{displaymath}
\langle M_s \rangle = T_H - {\dot T} \langle t \rangle , \; \sigma^2 =
{\dot T}^2 \langle t^2 \rangle +  T_H^2 -2 {\dot T}\langle t \rangle -
\langle M_s \rangle^2
\end{displaymath}
where the mean first-passage time $\langle t \rangle$ and $\langle t^2
\rangle$ are given  by ${\dot T} \langle t  \rangle = \int_{T_L}^{T_H}
e^{W(\theta)} d\theta$ and $ {\dot  T}^2 \langle t^2 \rangle = 2 \left
(  {\dot   T}  \langle  t   \rangle  T_H  -   \int_{T_L}^{T_H}  \theta
e^{W(\theta)} d\theta  \right )$. $W(\theta)$ is the  logarithm of the
probability  that   the  transition   has  not  occurred   before  the
temperature  $\theta$ is  reached,  and  is given  by  $W(\theta) =  -
\int_{\theta}^{T_H} \lambda \left ( u \right ) du / {\dot T}$.

In order to compare the predictions of the model with the experimental
results,  we  must  propose   a  function  $\lambda  (T)$.   Different
functions  were checked  and the  results obtained  were qualitatively
equivalent.  The simplest assumption is that $\lambda$ exhibits a pole
at  $T_L$ and  a zero  at  $T_H$, i.e.   $\lambda =  \omega (T_H-T)  /
(T-T_L)$, where $\omega$ is a characteristic frequency for nucleation.
By inserting  $\lambda$ into the expression for  $W(\theta)$ above, we
find:
\begin{displaymath}
W(\theta )=\frac{\omega}{\dot T} \left  [ (T_H-\theta) - (T_H-T_L) \ln
\left ( \frac{T_H - T_L}{\theta-T_L}\right ) \right ]
\end{displaymath}
By a simultaneous non-linear  minimum-$\chi^2$ fitting to the $\langle
M_s \rangle$  and $\sigma$ data  for Cu$_{68.4}$Al$_{27.8}$Ni$_{3.8}$,
we obtain the following estimations of the three-free-parameter model:
$T_H = 257 \pm  1$ K, $T_L= 242 \pm 1 $ K and  $\omega = (5.2 \pm 0.5)
10^{-2}$  s$^{-1}$.    The  fit   agrees  remarkably  well   with  the
experimental data  (Fig.  \ref{FIG4})  and reproduces the  increase of
$\sigma$ with $\dot  T$ and also the slightly  decreasing behaviour of
$\langle  M_s \rangle$.   The fitted  values  of $T_H$  and $T_L$  are
estimations of  the extreme values  of $M_s$.  Note that  $\omega (T_H
-T_L) \simeq 50 $ K/min renders an estimation of the cooling rate above
which     no     time      effects     will     be     observed     in
Cu$_{68.4}$Al$_{27.8}$Ni$_{3.8}$.    On  the   other  hand,   for  the
Cu$_{68.0}$Zn$_{16.0}$Al$_{16.0}$   sample  we  have   found  athermal
behaviour, even at 0.1 K/min.   Assuming that $\omega$ for this sample
is  similar to  the previous  one, $T_H-T_L  < 2  $K. $T_H$  and $T_L$
coincides  within  the  errors  and  no isothermal  behaviour  can  be
observed with the allowed temperature control resolution.

Our results  indicate that strictly, at  finite temperatures, athermal
transitions do not occur. However,  in practice, if $T_H-T_L$ is small
enough, kinetic effects are difficult to observe.  In this case (close
to the  athermal limit), scaling of  $(\nu / \dot T)$  versus $T$ will
occur for A.E.  measurements.  We have verified the robustness of such
scaling  in the  case of  the Cu-Zn-Al.   We claim  that,  in general,
$dx/dT = (dx/dt)/{\dot T}$ versus $T$ will be independent of $\dot T$,
where  $x$ is  the transformed  fraction. The  model proposed  in this
letter refers to the nucleation  of the first martensite domain in the
system.   The subsequent  growth  will be  characterised by  different
energy  barriers.  For  Cu-Zn-Al the  fact that  scaling  extends well
below the onset  of the transition, reveals that  the system satisfies
the conditions for it to behave athermally over all the transformation
range.  In contrast, scaling does not occur for Cu-Al-Ni either at the
transition onset or in the subsequent evolution.

A question still  remains open: why do the  two selected samples, with
transition  temperatures  within  the  same  temperature  range,  have
different $T_H  - T_L$ ranges?  We  argue that the origin  lies in the
small symmetry  differences between the  corresponding low temperature
phases:  orthorhombic and  hexagonal,  which give  rise to  different
accommodation   mechanisms  of   the  transformational   shape  change
necessary to minimise the elastic stored energy during the transition.
Microgliding  and  microtwinning   are  the  operative  mechanisms  in
Cu-Zn-Al  and Cu-Al-Ni  respectively \cite{Pelegrina90}.   Our results
are consistent with  the accepted idea that twinning  permits a larger
degree  of metastability.   

To  conclude, we  have proposed  a scaling  argument to  check  when a
first-order  phase  transition  can   be  treated  as  athermal.   The
different   behaviour   exhibited   by  martensites   with   different
transformation mechanisms  has been explained by a  single model based
on the mean first-passage time. In the light of present results, it is
clear that  for systems with  a lack of  scaling, the definition  of a
single transition  temperature $M_s$ is ambiguous: it  is a stochastic
variable  ranging  between $T_H$  and  $T_L$  which  are the  limiting
temperatures which characterize the transition.

We  would  like  to   thank  J.M.Sancho  and  A.Labarta  for  fruitful
discussions.   This work  has  received financial  support from  CICyT
(Project  No  MAT98-0315)  and  CIRIT (Project  2000SGR00025).   F.J.P
acknowledges financial support from DGICyT.
%



\begin{references}
\bibitem{porta} The  recorded number of  counts as a function  of time
has  been  double  checked  by  using  a  continuous  analog  counter,
connected  in  parallel  with  the  frequency  counter.   Results  are
perfectly comparable which  means that the dead times  inherent to the
gated frequency counter are irrelevant for the analysis.
\end{references}
\end{document}